\documentclass[preprint,5p,number,times,sort,compress]{elsarticle}

\usepackage{lineno}
\usepackage[colorlinks,
            linkcolor=blue, 
            anchorcolor=blue,
            citecolor=blue, 
            ]{hyperref}

\usepackage{color}
\usepackage{amsmath}
\modulolinenumbers[1]
\usepackage[utf8]{inputenc}
\usepackage{graphicx}
\usepackage{subcaption}
\usepackage{amsmath}
\usepackage{booktabs}

\usepackage{amssymb}
\usepackage{ulem}
\usepackage{multirow}
\usepackage{soul}
\setstcolor{red}

\usepackage{makecell}

\journal{EPJC}

\begin{document}
\begin{frontmatter}

\title{Implementation of a Low-Temperature Monitoring and Alarm System for the Taishan Neutrino Experiment}

\author[1]{Shengheng Huang}
\author[2]{Mei Ye\corref{cor1}}
\author[1]{Guihong Huang\corref{cor2}}
\author[2]{Yichen Li}
\author[2]{Zhimin Wang}
\author[3]{Xiaohao Yin}
\author[3]{Guang Luo}
\author[3]{Fengpeng An}
\author[4]{Jiahao Zhang}

\cortext[cor1]{Corresponding author. E-mail: yem@ihep.ac.cn}
\cortext[cor2]{Corresponding author. E-mail: huanggh@wyu.edu.cn}

\address[1]{Wuyi University, Jiangmen 529020, China}
\address[2]{Institute of High Energy Physics, Chinese Academy of Sciences, Beijing 100049, China}
\address[3]{Sun Yat-sen University, Guangzhou 510275, China}
\address[4]{Zhengzhou University, Zhengzhou 450001, China}

\begin{abstract}
The Taishan Antineutrino Observatory (TAO) is a near-site experiment for the Jiangmen Underground Neutrino Observatory (JUNO). Its primary goal is to provide a precise reference reactor antineutrino energy spectrum, thereby eliminating the model dependence in reactor neutrino spectrum predictions and enhancing the sensitivity of the neutrino mass ordering measurement. To ensure accurate data acquisition and safe operation of the TAO experiment, a low-temperature monitoring and alarm system has been developed. Built on the Experimental Physics and Industrial Control System (EPICS) framework, the system employs PT100 platinum resistance thermometers embedded in the detector to monitor the temperature of the liquid scintillator. Real-time temperature data are acquired, enabling comprehensive thermal monitoring. The alarm program adopts a trigger-based mechanism with multi-level thresholds, providing instant alerts to operators when the temperature deviates from the safe range. The system has been operating stably for six months, accumulating over one thousand alarm records, and has proven effective in ensuring the safe and stable operation of the experiment.
\end{abstract}


\end{frontmatter}

\section{Introduction}

The Jiangmen Underground Neutrino Observatory (JUNO) is the second large-scale neutrino experiment led by China, following the Daya Bay reactor neutrino experiment. Its central detector is a liquid scintillator detector featuring the highest energy resolution and the largest mass in the world. The primary scientific goal of JUNO is to determine the neutrino mass ordering using reactor neutrino oscillations~\cite{JUNO2022}.

Previous studies have revealed significant discrepancies between the measured reactor antineutrino spectrum and theoretical predictions~\cite{DayaBay2021}. To mitigate the impact of this uncertainty on the measurement of the neutrino mass ordering, the JUNO collaboration constructed a small, high-precision neutrino experiment at the Taishan nuclear power plant, approximately 30~m from the reactor core, known as the Taishan Antineutrino Observatory (TAO). TAO employs a low-temperature liquid scintillator to accurately measure the reactor antineutrino spectrum, providing essential input for JUNO~\cite{TAOCDR2020}.

The liquid scintillator in the TAO detector operates at \(-50\,^\circ\)C. To ensure safe and stable operation, real-time temperature monitoring and alarm capabilities are mandatory. In this work, we design and implement a distributed low-temperature monitoring and alarm system based on PT100 temperature sensors, a GM10 data acquisition system, and the Experimental Physics and Industrial Control System (EPICS). The system achieves the following functions:
\begin{enumerate}
    \item Temperature measurement accuracy better than \(\pm0.5\,^\circ\)C.
    \item Monitoring of the spatial uniformity of the temperature inside the detector.
    \item Multi-level alarms triggered in real time when temperature deviates from safety thresholds, allowing timely operator intervention.
    \item Archiving of temperature and alarm data into a database for subsequent physical analysis, and provision of a web-based remote monitoring interface.
\end{enumerate}
The system has been successfully deployed at TAO, providing critical support for thermal stability and detector performance.

\section{System Architecture of the Low-Temperature Monitoring and Alarm System}

\begin{figure*}[th]
\centering
\includegraphics[width=\linewidth]{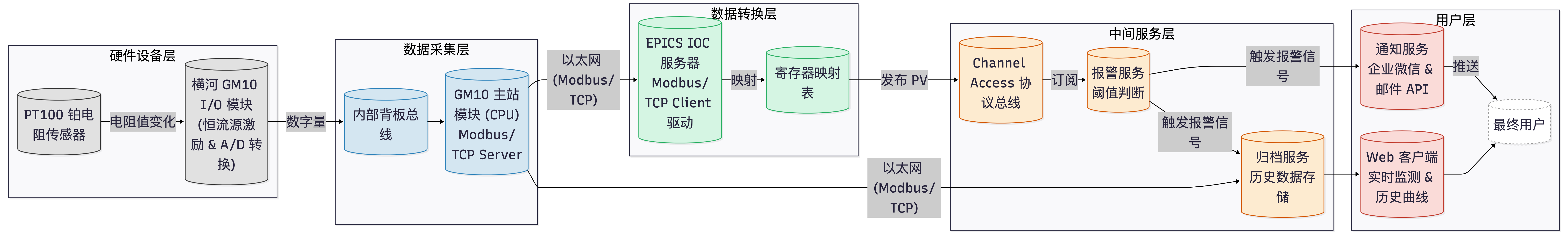}
\caption{Architecture diagram of the low-temperature monitoring and alarm system.}
\label{fig:architecture}
\end{figure*}

As a large-scale scientific facility, the TAO experiment requires its control system to maintain long-term stable operation, rapid response, and remote accessibility. For these reasons, EPICS is adopted as the core software framework. The overall architecture of the low-temperature monitoring and alarm system is shown in Fig.~\ref{fig:architecture}, consisting of five layers: hardware layer, data acquisition layer, data conversion layer, middleware layer, and user layer. This layered design implements complete monitoring and alarm functionality from the hardware to the user interface.

\begin{itemize}
    \item \textbf{Hardware layer:} PT100 platinum resistance temperature sensors are deployed in the cryogenic environment, converting temperature changes into resistance variations. The I/O modules of the Yokogawa GM10 system provide constant current excitation for the PT100s, convert the resistance signal into a voltage, and then digitize it via a high-precision A/D converter integrated into the module.
    \item \textbf{Data acquisition layer:} The GM10 I/O modules transmit the digitized temperature data to the GM10 master station CPU via the internal backplane bus. The master station then publishes the process data held in its registers using the Modbus/TCP protocol.
    \item \textbf{Data conversion layer:} The EPICS Input/Output Controller (IOC) reads the register data from the GM10 master station via a Modbus/TCP client driver and maps them to EPICS Process Variables (PVs).
    \item \textbf{Middleware layer:} The alarm system receives PV updates via CA callback mechanisms and performs anomaly detection based on preset thresholds. The archiving system persistently stores temperature and alarm data using PyMySQL.
    \item \textbf{User layer:} Users can view real-time temperature values, trend curves, and alarm status via a web client. When temperatures exceed safety thresholds, the system automatically pushes alarm notifications through WeChat Work (Enterprise WeChat) and email services.
\end{itemize}

\section{Hardware Design}

\subsection{Three-Wire Bridge Circuit Design}

Two-wire configurations suffer from significant measurement errors because the lead resistance adds directly to the sensor resistance, resulting in an error of approximately \(2.5\,^\circ\)C per ohm of lead resistance~\cite{Liu2020}. Four-wire configurations can achieve high precision (up to \(\pm0.02\,^\circ\)C) but at higher hardware cost~\cite{Zhang2002}. Over a lead resistance range of 0–20~\(\Omega\), three-wire configurations achieve an accuracy of about \(\pm0.1\) to \(\pm0.2\,^\circ\)C~\cite{Zhao2019}. Balancing accuracy and cost, our system adopts a three-wire bridge circuit design.

In this design, each PT100 sensor is connected to the measuring device via three wires: two for supplying the excitation current and one for measuring the voltage across the sensor. This configuration effectively compensates for errors caused by lead resistance, thus improving measurement accuracy~\cite{Gan2010}.

To monitor the spatial uniformity of the temperature inside the TAO detector, twenty PT100 probes are evenly distributed, as detailed in Section~\ref{sec:temperature-analysis}.

\subsection{Yokogawa GM10 Data Acquisition System}

\subsubsection{I/O Module Overview and Selection Justification}

The performance of different I/O modules is compared in Table~\ref{tab:io_modules}. The Yokogawa GM10 GX90XA-10-U2 module is a high-precision, multi-channel I/O module supporting 10 channels per module. Compared to the GX90XA-10-T1, the GX90XA-10-U2 offers better adaptability for three-wire PT100 measurements, faster scanning, and faster response~\cite{Li2024}. Compared to the GX90XA-06-R1, it provides more channels. Considering both channel count and performance, TAO selected two GX90XA-10-U2 modules to accommodate all twenty PT100 probes.

\begin{table*}[ht]
\centering
\caption{Performance comparison of different I/O modules.}
\label{tab:io_modules}
\begin{tabular}{lccc}
\toprule
\textbf{Comparison aspect} & \textbf{10-U2} & \textbf{06-R1} & \textbf{10-T1} \\
\midrule
Three-wire PT100 compatibility & Supported & Not supported & Supported \\
Number of PT100 channels & 10 & 6 & 10 \\
Scanning mode & Solid-state relay & Solid-state relay & Electromagnetic relay \\
Low-temperature response & 100~ms interval & 100~ms interval & 1~s interval \\
\bottomrule
\end{tabular}
\end{table*}

\subsubsection{EPICS-Based Integration of GM10 Data Acquisition}

To achieve unified access and distribution of temperature monitoring data within the EPICS control system, the data collected by GM10 must be integrated into the EPICS IOC via the Modbus/TCP protocol~\cite{Rivers2025}. This integration overcomes the previous limitation where GM10 data could only be accessed through the proprietary iDAQAnywhere R3 software or a limited web interface. After integration, all temperature channel readings are mapped in real time to standard EPICS PVs and managed within the IOC’s real-time database. Authorized clients (such as operator interfaces, alarm servers, and archiving engines) can then subscribe to and retrieve the real-time values and statuses of these PVs via the EPICS CA protocol.

\begin{figure}[ht]
\centering
\includegraphics[width=0.9\linewidth]{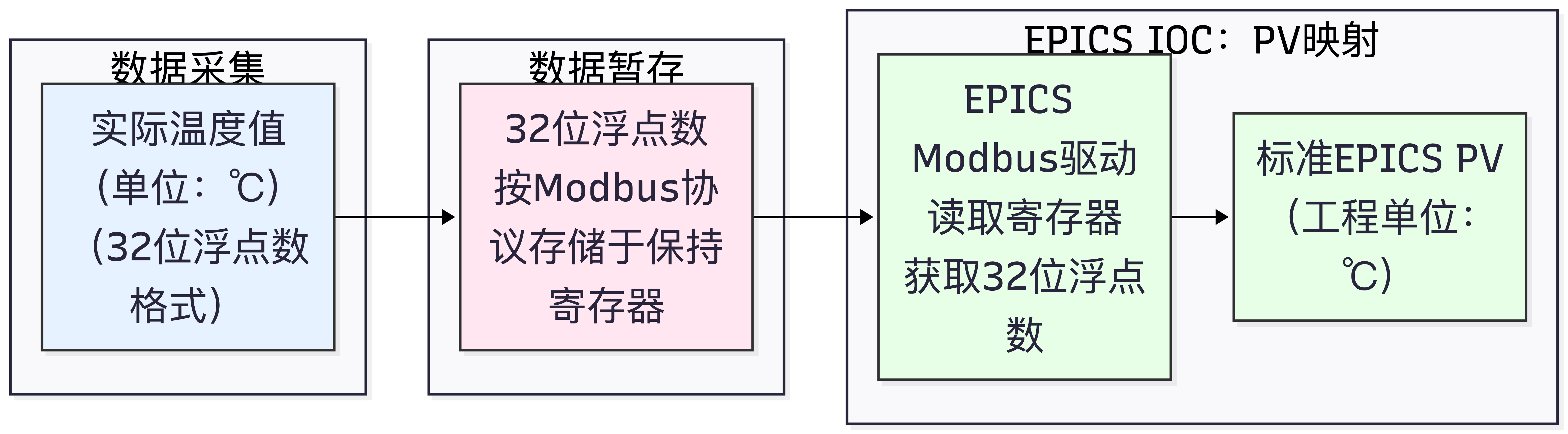}
\caption{Data conversion flow from GM10 to EPICS IOC.}
\label{fig:dataflow}
\end{figure}

Figure~\ref{fig:dataflow} illustrates the complete data conversion flow from the GM10 to the EPICS IOC~\cite{EPICSYokogawa}. First, the temperature data acquired by the GM10 are written to Modbus holding registers as 32-bit floating-point numbers, delivering the data from the acquisition front-end to the industrial protocol. Then, the Modbus/TCP driver loaded in the EPICS IOC reads the raw values from the predefined registers and converts them, based on a 1:1 linear mapping, into standard EPICS PVs with engineering units (\(^\circ\)C), alarm thresholds, and status attributes. For upper-layer applications, subscribing to a single PV via the CA protocol directly provides temperature data with timestamps and real-time status information.

The implementation steps are as follows:
\begin{enumerate}
    \item \textbf{Define register addresses:} The GM10 master station writes the digitized temperature values from each PT100 channel into its internal Modbus registers, each measurement point corresponding to a unique register address.
    \item \textbf{Configure GM10 network:} Assign a fixed IP address, subnet mask, and gateway to the GM10, ensuring it is on the same network segment as the IOC system, and enable the Modbus/TCP service.
    \item \textbf{IOC system configuration:} Configure the Modbus TCP client in the IOC to read GM10 data. This includes loading the Modbus device driver and registering the new device in the IOC development environment; setting device parameters (IP address of the GM10, default port 502); and defining PV variables, specifying their Modbus register addresses, data types (e.g., 32-bit float), and access attributes (e.g., read-only), ensuring strict consistency with the data format output by the GM10.
\end{enumerate}

\section{Software Design}

\subsection{Alarm Logic}

The alarm program employs a trigger-based monitoring mechanism to monitor the liquid scintillator temperature in real time, notifying operators promptly when anomalies occur. Each PV can be in one of three states: normal (status code 0), level-1 alarm (status code 1), or level-2 alarm (status code 2). The state is determined by whether the PV value exceeds preset thresholds. When the status code jumps from a lower level to a higher level, the system triggers the corresponding alarm.

\begin{figure}[htbp]
\centering
\includegraphics[width=0.9\linewidth]{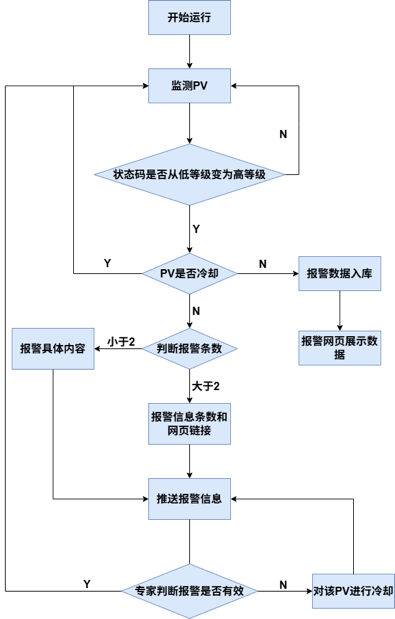}
\caption{Flowchart of the alarm program logic.}
\label{fig:alarm_logic}
\end{figure}

The logic flow of the alarm program is shown in Fig.~\ref{fig:alarm_logic}. When the status code of a PV increases, the system determines that an anomaly has occurred at that measurement point and initiates an exception-handling process: it first checks whether the region corresponding to the PV is in the cooling phase (i.e., satisfies a predefined alarm masking condition). If yes, it returns to monitoring the PV. Otherwise, it checks whether the number of active alarms exceeds two. If so, it pushes a summary message (including the alarm count and a link to the alarm webpage); otherwise, it pushes detailed alarm content. If the operator confirms that an alarm is redundant or false, the PV can be set to a cooling state to suppress repeated alarms. Subsequently, the system stores the complete alarm data in the database, pushes a notification via WeChat Work, and simultaneously displays the relevant data on the web monitoring interface for the on-duty operator.

\subsubsection{Data Acquisition}

Data acquisition for EPICS PVs follows a producer-consumer architecture based on the CA protocol’s subscription-notification mechanism, which is event-driven and requires no active polling.

When the program creates an EPICS PV object, it registers a callback function: \\\texttt{on\_change} via \texttt{pv.add\_callback(self.on\_change)}. This PV object is managed by the EPICS CA library at a low level, which starts a dedicated thread to continuously listen for data updates from the server. Whenever a PV value changes, the CA library receives the update over the network and triggers the callback. In this model, the \texttt{on\_change} callback acts as the producer, running in the CA library thread. It receives PV updates, performs initial status determination, and writes the alarm data to be processed into a shared queue. The functions \texttt{process\_raw\_alarms} and \texttt{send\_message\_queue} act as consumers, reading data from the queue and performing deduplication, summarization, and distribution of alarm messages. This design effectively decouples data acquisition from business processing, enhancing system real-time performance and stability.

\subsubsection{Alarm Level Design}

Two alarm levels are defined: general alarm and severe alarm, corresponding to different temperature thresholds.
\begin{itemize}
    \item \textbf{General alarm:} Temperature deviation \(\pm0.5\,^\circ\)C from the setpoint.
    \item \textbf{Severe alarm:} Temperature deviation \(\pm1.0\,^\circ\)C from the setpoint.
\end{itemize}
The specific threshold settings are listed in Table~\ref{tab:alarm_levels}.

\begin{table}[ht]
\centering
\caption{Alarm level design. The threshold are customizable.}
\label{tab:alarm_levels}
\begin{tabular}{lcc}
\toprule
\textbf{Alarm level} & \textbf{Alarm type} & \textbf{Threshold} \([^\circ C]\) \\
\midrule
-1 level & General too low & Temperature < \(-50.5\) \\
-2 level & Severe too low & Temperature < \(-51.0\) \\
1 level & General too high & Temperature > \(-49.5\) \\
2 level & Severe too high & Temperature > \(-49.0\) \\
\bottomrule
\end{tabular}
\end{table}

\subsubsection{Alarm Uniqueness}

Alarm uniqueness determines whether multiple alarm signals correspond to the same abnormal event. For multiple triggered alarms, factors such as measurement reliability, false positive rate, and correlation with the anomalous event must be considered to select a single representative effective alarm. Improper uniqueness determination may lead to distorted or redundant alarm information. The strategies adopted by this system are detailed in Table~\ref{tab:alarm_uniqueness}.

\begin{table}[ht]
\centering
\caption{Alarm uniqueness issues and solution strategies.}
\label{tab:alarm_uniqueness}
\begin{tabular}{p{0.2\linewidth}p{0.3\linewidth}p{0.35\linewidth}}
\toprule
\textbf{Issue} & \textbf{Definition} & \textbf{Solution} \\
\midrule
Persistent alarm & Temperature exceeds the threshold for a long time; the system continues to identify the PV as in alarm. & Set alarm trigger condition to a PV status code increase. \\
Repeated alarm & Temperature fluctuates near the threshold, causing repeated alarms from the same PV. & Apply a cooling mechanism to alarm content from the same PV. \\
\bottomrule
\end{tabular}
\end{table}

\subsubsection{Alarm Message Push}

\begin{figure}[ht]
\centering
\begin{subfigure}[b]{0.4\linewidth}
    \centering
    \includegraphics[width=\linewidth]{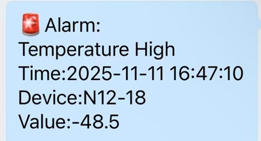}
    \caption{Single alarm}
    \label{fig:single_alarm}
\end{subfigure}
\begin{subfigure}[b]{0.45\linewidth}
    \centering
    \includegraphics[width=\linewidth]{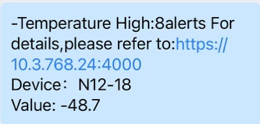}
    \caption{Multiple alarms}
    \label{fig:multi_alarm}
\end{subfigure}
\caption{Alarm push formats.}
\label{fig:alarm_push}
\end{figure}

Alarm messages are pushed via two channels: WeChat Work (Enterprise WeChat) and email, primarily targeting the management personnel. WeChat Work is centrally managed by the Computing Center of the Institute of High Energy Physics. Email push supports IHEP mailboxes, NetEase mailboxes, and QQ mailboxes. The push strategies are as follows:
\begin{enumerate}
    \item If only a single alarm is generated within a given time window, the alarm information is integrated and pushed directly, including the alarm type, time, and specific value (see Fig.~\ref{fig:single_alarm}).
    \item If multiple alarms are generated within the time window, a summary of all alarms is pushed, including the total number of alarms and a link to the alarm monitoring webpage for further query (see Fig.~\ref{fig:multi_alarm}).
\end{enumerate}

\begin{figure}[ht]
\centering
\includegraphics[width=\linewidth]{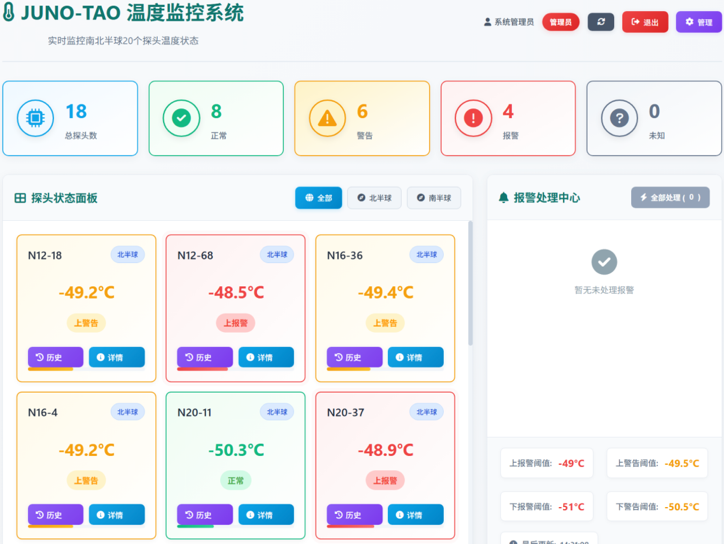}
\caption{Alarm webpage main page.}
\label{fig:main_page}
\end{figure}

\begin{figure}[ht]
\centering
\includegraphics[width=\linewidth]{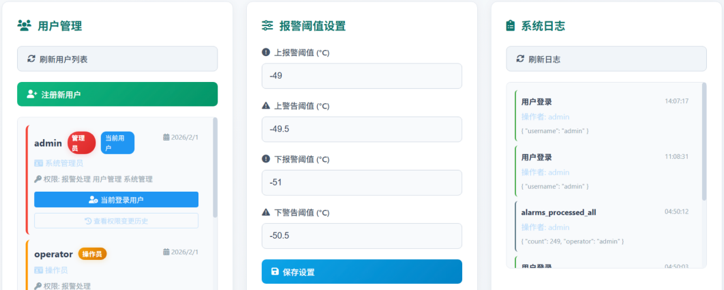}
\caption{Alarm webpage administration page.}
\label{fig:admin_page}
\end{figure}

\begin{figure}[ht]
\centering
\includegraphics[width=\linewidth]{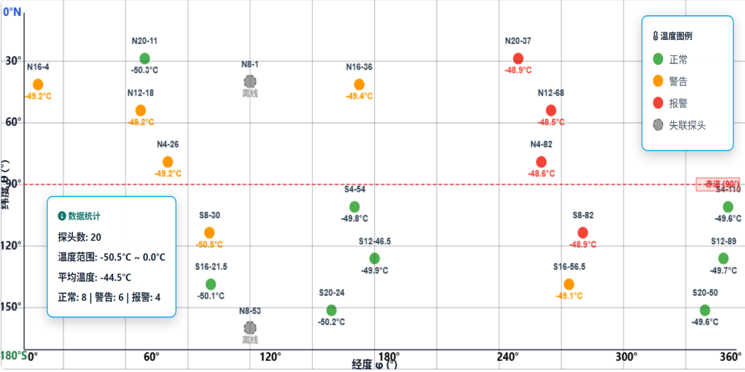}
\caption{Real-time spatial temperature distribution map of the TAO detector.}
\label{fig:spatial_distribution}
\end{figure}

\subsection{Alarm Webpage}

The alarm webpage is developed using HTML, CSS, and JavaScript front-end technologies, and consists of three parts: a login page, a main page, and an administration page.

The login page allows the administrator to create user accounts and assign one of three roles: administrator, operator, or observer. The administrator has full operation permissions, including viewing operation logs, handling alarms, adjusting alarm thresholds, and managing user accounts and roles. Operators are limited to alarm handling. Observers can only view real-time temperature data.

The main page (Fig.~\ref{fig:main_page}) displays real-time temperature data from all probes, with current alarm information shown on the right. After an alarm is triggered, operators and administrators can handle it. When the number of alarms is large, a one-click handling function is provided. Each probe area features two buttons: “History” and “Details”. The “History” button queries the last ten alarm records for that probe, and the “Details” button shows the probe status and channel configuration information.

The administration page (Fig.~\ref{fig:admin_page}) is accessible only to the administrator. Its main functions include monitoring system operation status (such as the total number of registered users and permission distribution), managing user accounts (including adjusting permissions and removing long-term inactive accounts), setting alarm and warning thresholds, and querying all login and operation logs to ensure system security and traceability.

The temperature monitoring system deploys ten temperature probes in the northern hemisphere and ten in the southern hemisphere, almost uniformly distributed on the outer surface of the TAO detector’s copper shell. Figure~\ref{fig:spatial_distribution} presents a spatial temperature distribution map generated from real-time probe data, used to monitor the temperature uniformity inside the detector. This map is deployed on the alarm webpage. On this map, a green circle indicates normal temperature, orange indicates a warning, red indicates an alarm, and gray indicates a lost connection.

\section{Temperature Data Analysis}
\label{sec:temperature-analysis}

\begin{figure}[ht]
\centering
\includegraphics[width=0.7\linewidth]{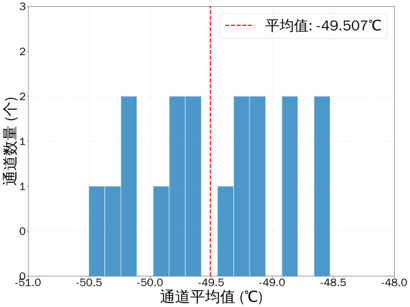}
\caption{Distribution of mean temperatures of TAO detector probes.}
\label{fig:temp_mean}
\end{figure}

\begin{figure}[ht]
\centering
\includegraphics[width=0.7\linewidth]{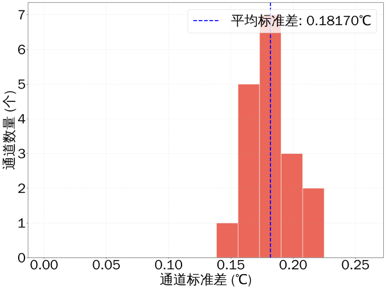}
\caption{Distribution of temperature standard deviations of TAO detector probes.}
\label{fig:temp_std}
\end{figure}

\begin{figure}[ht]
\centering
\includegraphics[width=\linewidth]{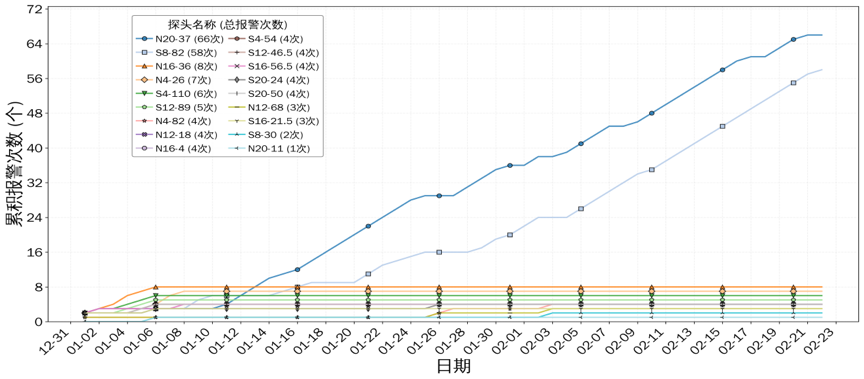}
\caption{Cumulative alarm count over time (18 channels).}
\label{fig:alarm_accum}
\end{figure}

\begin{figure}[ht]
\centering
\includegraphics[width=\linewidth]{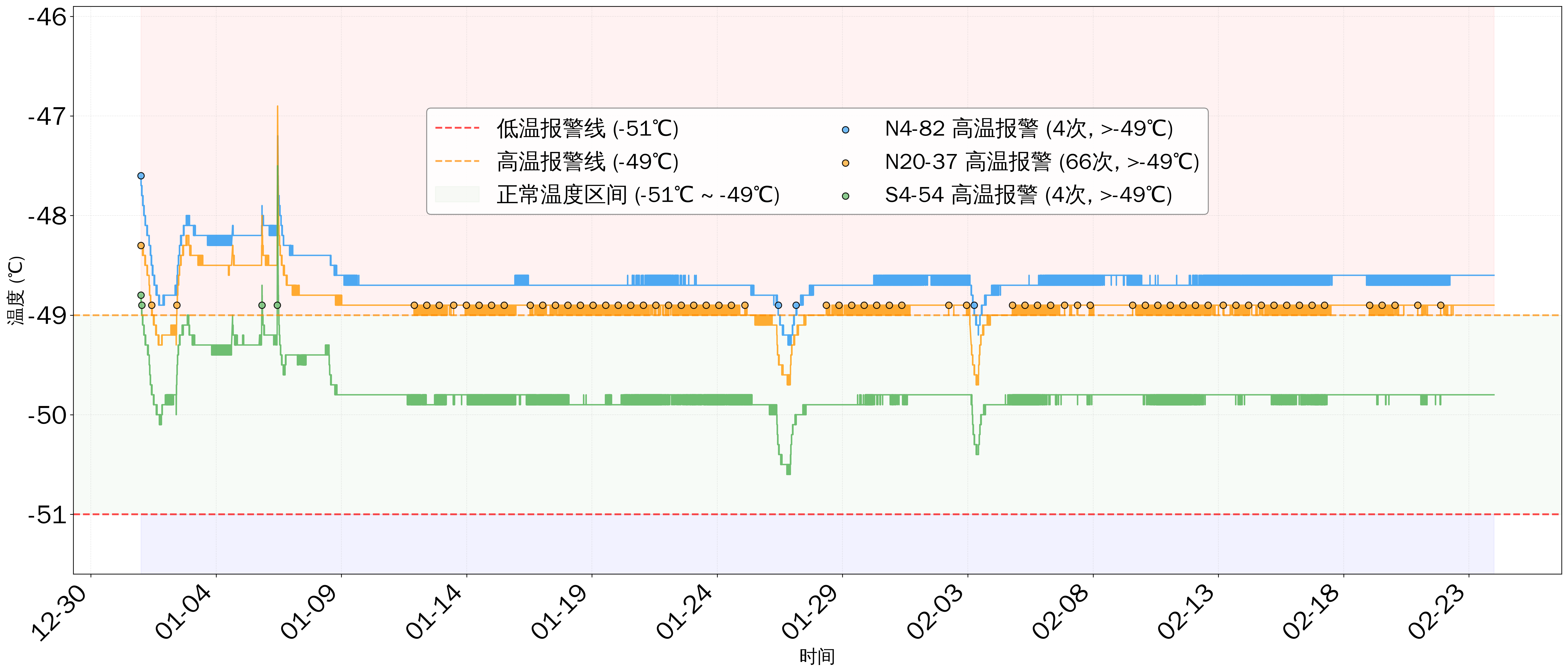}
\caption{ Long‑term temperature variation curves of three representative cold‑temperature probes. The red dashed line and orange dashed line denote the low‑temperature alarm threshold ($-51^\circ\text{C}$) and high‑temperature alarm threshold ($-49^\circ\text{C}$), respectively. The red dots mark high‑temperature alarm events triggered by probe N4‑82, N20‑37 and S4‑54, with corresponding alarm counts indicated in the legend.}
\label{fig:TvsT}
\end{figure}

\begin{figure}[ht]
\centering
\includegraphics[width=\linewidth]{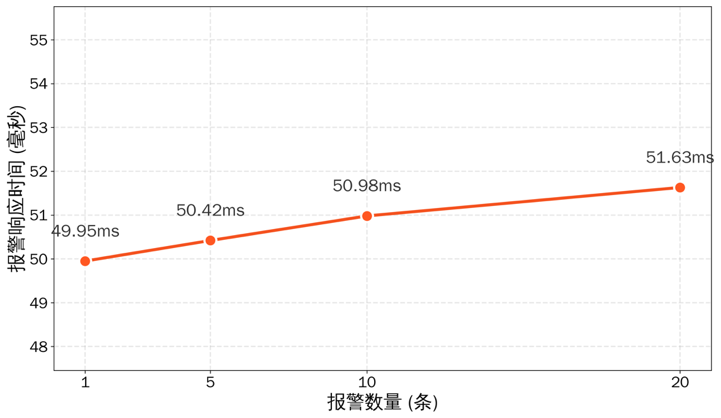}
\caption{Alarm response time versus number of concurrent alarms.}
\label{fig:response_time}
\end{figure}

The TAO detector is now in full operation. Based on monitoring data from January 1 to February 22, 2026 (53 consecutive days), we performed a statistical analysis of the temperatures recorded by 20 low-temperature probes (including two faulty ones). The temperature acquisition period is set to one minute. The results are shown in Figs.~\ref{fig:temp_mean} to \ref{fig:response_time}.

Figure~\ref{fig:temp_mean} shows a histogram of the mean temperature of each probe. The horizontal axis represents the mean temperature of a probe, and the vertical axis represents the number of probes falling into each interval. The red dashed line indicates the average of the mean temperatures across all probes. The results show that the mean temperatures of 14 probes are concentrated in the range \(-50.5\,^\circ\)C to \(-49.0\,^\circ\)C, while the mean temperatures of the remaining 4 probes are above the severe high-temperature alarm threshold (\(-49.0\,^\circ\)C).

Figure~\ref{fig:temp_std} shows a histogram of the standard deviation of each probe’s temperature data. The horizontal axis represents the standard deviation (reflecting temperature fluctuations), and the vertical axis represents the number of probes in each interval. The blue dashed line indicates the average standard deviation. The standard deviations of the probes are mainly distributed between \(0.15\,^\circ\)C and \(0.25\,^\circ\)C, indicating good measurement stability.

Figure~\ref{fig:alarm_accum} summarizes the accumulated alarm records of the low-temperature probes over the nearly two-month period. The cooling mechanism masks alarms from the same probe with the same temperature value within 12 hours. During the statistical period, the 18 effective probes triggered a total of 191 level-2 alarms: 97 alarms occurred in the northern hemisphere (8 probes) and 94 in the southern hemisphere (10 probes). The highest number of alarms for a single probe was 66, recorded at the northern hemisphere probe N20-37, because this probe was persistently in a slightly overheated state (see Fig.~\ref{fig:TvsT}). This alarm distribution provides data support for future system reliability optimization and cooling mechanism refinement.

To evaluate alarm performance, response time tests were conducted. By adjusting the alarm thresholds, four load levels ranging from 1 to 20 concurrent PV alarms were set. The alarm response time is defined as the time difference between the alarm push and the PV state change. Figure~\ref{fig:response_time} presents the test results: the horizontal axis is the number of concurrent alarms, and the vertical axis is the response time. The results show that although the response time tends to increase with the number of concurrent alarms, it remains below 52 ms, demonstrating excellent real-time performance.

\section{Conclusion}

This work has developed a 20-channel distributed temperature monitoring and alarm system based on PT100 temperature sensors, a Yokogawa GM10 data acquisition system, and EPICS, together with a multi-level trigger-based alarm program. After more than six months of continuous operation, the system demonstrates the following performance and value:
\begin{enumerate}
    \item \textbf{High accuracy and stability:} Temperature measurement accuracy is better than \(\pm0.5\,^\circ\)C, and the standard deviation of long-term temperature fluctuations is controlled between \(0.15\) and \(0.25\,^\circ\)C, meeting the stringent low-temperature requirements of the TAO experiment.
    \item \textbf{High reliability:} No software crashes or data interruptions were recorded during the 53-day statistical period, and more than one thousand valid alarms were processed. The “state-transition trigger + cooling mechanism” effectively suppresses repeated alarms, ensuring that operators focus on genuine anomalies.
    \item \textbf{High real-time performance:} Under the extreme scenario of 20 concurrent alarms, the alarm push delay is less than 52 ms, fully satisfying the real-time standards for industrial-grade monitoring.
\end{enumerate}
The system has been fully validated in the TAO experiment and provides a cost-effective, replicable reference model for the construction of monitoring systems in small- to medium-scale physics experimental setups. Future work will focus on introducing machine learning methods for predictive alarm based on temperature trends, further enhancing the level of intelligent operation and maintenance.

\section*{Acknowledgments}

This work was supported by the National Natural Science Foundation of China (Grant No. 12405231), the Characteristic Innovation Project for Regular Higher Education Institutions of Guangdong Provincial Department of Education (2024KTSCX044), and the Science Foundation for High-Level Talents of Wuyi University (2021AL027).

\end{document}